\documentstyle[12pt,aaspp4]{article}
\received{}
\accepted{}

\lefthead{Basu}
\righthead{Circumstellar Disks}

\begin{document}

\title{Constraints on the Formation and Evolution of 
Circumstellar Disks in Rotating Magnetized Cloud Cores}

\vspace{0.2in}

\author{\sc Shantanu Basu}
\affil{Canadian Institute for Theoretical Astrophysics,
University of Toronto, \\60 St. George Street, Toronto, Ontario M5S 3H8, 
Canada; basu@cita.utoronto.ca.}
\authoremail{basu@cita.utoronto.ca}

\vspace{0.6in}

\newcommand{\bl}[1]{\mbox{\boldmath$ #1 $}}
\newcommand{\Bref}{B_{\rm ref}}
\newcommand{\Omb}{\Omega_{\rm b}}
\newcommand{\cs}{c_{\rm s}}
\newcommand{\rc}{r_{\rm c}}
\newcommand{\rd}{r_{\rm d}}
\newcommand{\ms}{m_{\rm s}}
\newcommand{\md}{m_{\rm d}}
\newcommand{\ksid}{k_{\rm SID}}
\newcommand{\ksimd}{k_{\rm SIMD}}
\newcommand{\ampbyt}{a_{\rm M,P}/a_{\rm T}}
\newcommand{\amtbyg}{a_{\rm M,T}/g_r}
\newcommand{\muG}{\mu{\rm G}}
\newcommand{\tni}{\tau_{\rm ni}}
\newcommand{\vnr}{v_{{\rm n},r}}
\newcommand{\beq}{\begin{equation}}
\newcommand{\eeq}{\end{equation}}
\renewcommand{\and}{\&\ }
\newcommand{\TM}{Mouschovias, T. Ch.}
\newcommand{\refer}{\reference}

\begin{abstract}
We use magnetic collapse models to place some constraints on the
formation and angular momentum evolution of circumstellar disks
which are embedded in magnetized cloud cores. Previous models have
shown that the early evolution of a magnetized cloud core is governed
by ambipolar diffusion and magnetic braking, and that the core
takes the form of a nonequilibrium flattened envelope
which ultimately collapses dynamically to form a protostar. In this paper, 
we focus on the inner centrifugally-supported disk, which
is formed only after a central protostar exists, and grows by dynamical
accretion from the flattened envelope. We estimate a centrifugal radius for
the collapse of mass shells within a rotating, magnetized cloud core.
The centrifugal radius of the inner disk is related to its mass through the
two important parameters characterizing the background medium:
the background rotation rate $\Omb$ and the background
magnetic field strength $\Bref$. 
We also revisit the issue of how rapidly mass is deposited onto the disk
(the mass accretion rate) and use several recent models to comment upon
the likely outcome in magnetized cores.
Our model predicts that a significant centrifugal disk (much larger than a
stellar radius) will be present in the very early (Class 0) stage of
protostellar evolution.
Additionally, we derive an upper limit for the disk radius as it evolves
due to internal torques, under the
assumption that the star-disk system conserves its mass and angular momentum
even while most of the mass is transferred to a central star.
\end{abstract}

\keywords{accretion, accretion disks - circumstellar matter - ISM: clouds 
- ISM: magnetic fields - MHD - stars: formation - stars: pre-main-sequence}

\section{INTRODUCTION}

Ever since the discovery that the planets in our solar system have
nearly circular and coplanar orbits, the presence of a gaseous circumstellar
disk has been thought to be intimately associated with star and
planet formation. At the present time, circumstellar disks can be detected
around a large number of young stellar objects (YSO's) by a variety
of techniques, including
infrared excess (e.g., Cohen, Emerson, \& Beichman 1989; Strom et al. 1989), 
measurement of their millimeter and submillimeter
flux (e.g., Beckwith et al. 1990; Adams, Emerson, \& Fuller 1990; 
Ohashi et al. 1991; Osterloh \& Beckwith 1995), and even direct optical
imaging of disk silhouettes in the Orion Nebula using HST
(McCaughrean \& O'Dell 1996). The millimeter and submillimeter measurements, 
which are sensitive to the outer parts of the disk, usually find disk
radii $\lesssim 100$ AU. The radii of the Orion silhouettes range from
25 AU to 500 AU. Altogether, a disk radius of $\sim 100$ AU is often taken 
to be a typical value. However, circumstellar disks of size $\sim 1000$ AU
are also found around some older stars such as $\beta$ Pictoris.  

Theoretically, disks are believed to be a natural consequence of the
collapse of a rotating molecular cloud core.
Therefore, in principle, one should be able to derive some physical
properties of a circumstellar disk directly from the properties of a cloud
core in which it is formed, or even from the properties of the ambient
molecular cloud in which the core itself was formed.
A crucial quantity that should be derived from a collapse model
is the centrifugal radius for infalling mass shells; this quantity
depends certainly on the existing density and angular velocity profile
near a protostar at the time that it is formed. Previous estimates
of the centrifugal radius have tended to use hydrodynamic models.
For example, Terebey, Shu, \& Cassen (1984) have derived a centrifugal
radius $\rc$ under the assumption that the core is uniformly rotating
and has the static density profile of a singular isothermal sphere at the
moment that a central protostar is formed.

In this paper, we estimate a centrifugal radius for the collapse of a rotating,
magnetized core. We use results from numerical magnetohydrodynamic (MHD)
models of the formation of cores by ambipolar diffusion and magnetic braking
(Basu \& Mouschovias 1994, 1995a, 1995b, hereafter BM94, BM95a,b) and their
subsequent collapse once they have a supercritical mass-to-flux ratio.
The nonrotating aspects of these MHD models are discussed in detail by
Fiedler \& Mouschovias (1992, 1993, hereafter FM93) and Ciolek \& Mouschovias
(1993, 1994, 1995, hereafter CM93, CM94, CM95). In a previous paper,
(Basu 1997), we used these numerical models to build a
semianalytic model for the supercritical collapse phase.
Our analytic expressions for the inner solution could be extrapolated
to the instant that
a central protostar is formed. At this moment, labeled $t=0$ in the
terminology of isothermal similarity solutions for cloud collapse
(Larson 1969; Shu 1977; Hunter 1977), the cloud core takes
the form of a flattened envelope around the protostar, with a
nonequilibrium power-law radial density profile ($\rho \propto r^{-2}$)
and supersonic infall velocity. The core is also differentially rotating,
with angular velocity $\Omega \propto r^{-1}$ in the innermost region.
Although the semianalytic solution applies to the time period $t < 0$,
we can extrapolate to $t>0$ for the purpose
of estimating the centrifugal radius, since angular momentum should
continue to be conserved in this more dynamical phase of
accretion onto the central protostar. A centrifugal radius
is expected to exist during the accretion phase ($t>0$), even though it does
not exist during the earlier runaway collapse phase ($t<0$).

We relate the centrifugal radius $\rc$
to the disk mass $m$ though two fundamental parameters characterizing
the ambient cloud:
the background rotation rate $\Omb$, and the background magnetic field
strength $\Bref$. The mass accretion rate $\dot{m}$ is dependent on two other
fundamental parameters: the
isothermal sound speed $\cs$ and the universal gravitational constant $G$.
However, $\dot{m}$ is not expected to be constant for
$t>0$, for reasons given by Basu (1997). 
This has been demonstrated by recent numerical models of magnetized clouds
(Tomisaka 1996; Ciolek \& K\"onigl 1998).
We discuss the implications of several recent papers (Basu 1997; 
Saigo \& Hanawa 1998; Ciolek \& K\"onigl 1998; Tomisaka 1996) on the mass
accretion rate, and use them to place constraints on its likely value.

Once formed, or perhaps during the formation period itself, the disk
is expected to undergo significant angular momentum evolution
in order to give birth to star-disk systems of the type that are observed.
Given the current theoretical uncertainties in understanding these processes,
we present a constraint on the disk evolution if driven by internal torques
(e.g., viscous or gravitational). Regardless of the details of the process,
we find an upper limit to the disk radius if its
mass and angular momentum are conserved.

We also discuss how estimates of disk sizes in the early protostellar
phases (before the protostar has assembled most of its mass) may help
to distinguish between various collapse scenarios which yield different
physical properties in the near-protostellar environment.
In essence, we find that a differentially rotating nonequilibrium 
pre-stellar core is necessary in order to obtain a significant centrifugal
disk in the early (Class 0) protostellar phase.

\section{Formation of an Inner Centrifugal Disk}

\subsection{The Centrifugal Radius}

In the presence of dynamically significant magnetic fields,
star formation is expected to occur within a flattened cloud core
aligned perpendicular to the mean magnetic field direction. 
FM93 found that an initially magnetically
subcritical (i.e., mass-to-flux ratio below the critical value for
collapse) cloud will first relax to equilibrium along field lines
as it begins a quasistatic, ambipolar diffusion driven radial
contraction toward a local density peak. This radial contraction becomes
dynamic within a central region after it has achieved a supercritical
mass-to-flux ratio; however, near-equilibrium is maintained along
field lines. The supercritical regions, which are identified with
molecular cloud cores, provide a flattened envelope within which 
stars form and grow by accretion. These disklike structures are
in fact much larger (size up to $\sim 10,000$ AU) than a possible inner 
centrifugally-supported disk, and are not in equilibrium
(they are supercritical). They are also not to 
be confused with the ``pseudodisks'' of Galli \& Shu (1993a,b), which form 
{\it after} the formation of a central protostar in a cloud that 
has the density profile of a 
singular isothermal sphere when the protostar is formed, and is 
threaded by a dynamically weak magnetic field. When magnetic fields
are dynamically significant, as implied by some Zeeman measurements 
(Goodman et al. 1989; Crutcher et al. 1993), the formation of the outer
magnetic disk {\it precedes} star formation. 

The collapse of magnetic cores has been further investigated numerically by 
CM94, 95, and BM94, 95a, 95b, using the thin-disk approximation. Basu (1997)
found analytic expressions for the inner profiles of physical variables 
in these collapsing cores. A natural limiting form for isothermal
collapse is when a power-law density profile (established due to 
self-similarity in the innermost region) extends inward to radius $r=0$,
creating a central singularity. Since a finite mass now exists at $r=0$,
this moment is usually associated physically with the formation of a
central protostar.\footnote{Isothermality is not really expected to  
be valid all the way to $r=0$, i.e., for all densities, but it is a 
reasonable first step to pursue, since the non-isothermal gas occupies
an inner region of size $\lesssim 10$ AU (Larson 1969).} Basu (1997) shows that
the magnetic collapse models of BM94, 95a, 95b typically lead to the limiting 
column density (integrated along the mean magnetic field direction) profile
\beq
\sigma(r) \simeq \frac{\cs^2}{Gr} \label{sig}
\eeq
where $\cs$ is the isothermal sound speed and $G$ is the universal
gravitational constant. The above equation describes a {\it nonequilibrium}
profile (see \S 2.2). 
The angular velocity, which becomes aligned with the mean magnetic field 
direction during the core formation epoch (Mouschovias \& Paleologou
1979, 1980), achieves the limiting profile
\beq
\Omega(r) \simeq \frac{2 \pi \Omb \cs^2}{\Bref G^{1/2} r} \label{omega}
\eeq
(Basu 1997), where $\Omb$ is the ambient rotation rate of the cloud and
$\Bref$ is a uniform background (or ``reference'') magnetic field.
This differential rotation profile is a result of magnetic braking
enforcing uniform rotation $\Omega \simeq \Omb$ until a critical 
column density is reached, and subsequent near angular momentum 
conservation during the rapid collapse phase, during which the column density
profile of equation (\ref{sig}) is built up (BM94). Based on the BM94
results, Basu (1997) showed that
angular momentum conservation effectively begins at a column density
$\sigma_{\rm crit} \simeq 2 \Bref/\mu_{\rm crit}$, where 
$\mu_{\rm crit} = (2 \pi G^{1/2})^{-1}$ is the critical mass-to-flux 
ratio for a magnetized thin disk (Nakano \& Nakamura 1978). 
The result is a rotation rate many orders of magnitude greater than 
$\Omb$ near the cloud center.
Equations (\ref{sig}) and (\ref{omega}) yield the following relation between
specific angular momentum $j=\Omega\,r^2$ and the enclosed mass $m$:
\beq
j \simeq \frac{\Omb G^{1/2}}{\Bref}\, m. \label{jm}
\eeq
A typical (mean) rotation rate for dense molecular cloud cores or starless
Bok globules is about $10^{-14}$ rad s$^{-1}$ (Goodman et al. 1993;
Kane \& Clemens 1997). Using this as a typical value for the pre-collapse
rotation rate $\Omb$, and using a canonical magnetic field value of
$30 \muG$ (see Goodman et al. 1989; Crutcher et al. 1993), equation
(\ref{jm}) yields $j \simeq 10^{20} - 10^{21}$ cm$^2$ s$^{-1}$ for
$m \simeq 1 - 10$ M$_{\sun}$. The upper range is in agreement with 
estimates for $j$ in dense cores or Bok globules (Goodman et al. 1993;
Kane \& Clemens 1997), and the lower range is in good agreement with 
estimates of $j$ for smaller scale protostellar envelopes 
(Ohashi et al. 1997b), for which the lower mass range may be more relevant.

The $j-m$ relation of equation (\ref{jm}) is preserved during 
the dynamic collapse phase before
a central protostar is formed ($t < 0$), since the collapse timescale
becomes much shorter than the magnetic braking timescale (BM94). 
An inner centrifugally-supported disk may be expected to form near the
cloud center, where the rotation rate is the highest. However,
during the time period $t < 0$, the ratio of centrifugal acceleration
to gravitational acceleration (the centrifugal support) remains 
constant in the central region, i.e., both accelerations increase
in proportion to $r^{-3}$, where $r$ is a Lagrangian radius.
This is a property of the self-similar collapse, and occurs in a rotating,
magnetic disk (BM95a; Basu 1997), where efficient magnetic braking prior
to collapse keeps the centrifugal support at very low levels anyway.
This effect is also present in
non-magnetic rotating disks (Norman et al. 1980;
Narita, Hayashi, \& Miyama 1984; Hayashi 1987), where the centrifugal
support is much greater but also cannot grow and thereby halt the collapse.
After a central protostar is formed ($t>0$), the collapse becomes even
more dynamic in an inner region where the infall resembles free-fall
onto a central point mass, e.g., the similarity solutions of
Shu (1977) and Hunter (1977). Hence, angular momentum is more likely
to be conserved during this phase. More importantly, the nature of 
self-gravity in the flattened envelope is different at this stage, 
particularly near the cloud center.
At $t=0$, an $r^{-1}$ column density profile yields a gravitational field
$g_r = - Gm/r^2$, where $m$ is the enclosed mass, i.e., the same as in
spherical geometry. For example, equation (\ref{sig}) yields
$m(r) = 2 \pi \cs^2 r/G$, so that
the integral equation for a thin-disk gravitational field (see equations
[12] and [13] of BM94) yields $g_r = - 2 \pi \cs^2/r = - Gm/r^2$.
When $t>0$, the inner near free-fall region has infall velocity
$v_r \propto r^{-1/2}$, and column density $\sigma \propto r^{-1/2}$ 
if the mass accretion rate $\dot{M} = - 2 \pi \sigma r v_r$
is uniform in this region\footnote{
A uniform mass accretion rate 
is a property of self-similar collapse and results in the
inner density profile $\rho \propto r^{-3/2}$ in spherical
similarity solutions (e.g., Shu 1977; Hunter 1977). It also results in
an inner column density profile $\sigma \propto r^{-1/2}$ 
in thin-disk similarity solutions, as first noted by Nakamura, Hanawa, 
\& Nakano (1995).}.
The $\sigma \propto r^{-1/2}$ profile does not yield the exact
relation $g_r = - Gm/r^2$, but numerical integration shows that
$g_r = - g_0 \, Gm/r^2$, where $g_0 \simeq 0.7$ (Ciolek \& K\"onigl 1998).
Hence, during $t \geq 0$, when the column density profile is a combination
of $r^{-1/2}$ and $r^{-1}$, we expect $g_r = - g_0 \,Gm/r^2$,
with $g_0$ falling  in the range $0.7 - 1$. 
For the remainder of this paper, we simply use
$g_r \simeq  - Gm/r^2$, and not consider the
multiplicative constant in front that is close to unity.

Since the gravitational field increases
less rapidly than the centrifugal acceleration during $t>0$, each mass shell
(of fixed $j$) hits a centrifugal barrier at a radius $\rc$ determined
(to within a factor of order unity) by the relation
\beq
\frac{j^2}{\rc^3} \simeq \frac{Gm}{\rc^2}.
\eeq
Incorporating equation (\ref{jm}), we find the centrifugal radius 
\beq
\rc \simeq  \left(\frac{\Omb}{\Bref}\right)^2 m. \label{rc}
\eeq
Hence, the centrifugal radius can be related to the disk mass via two
fundamental parameters characterizing the ambient molecular cloud.
Using canonical values for molecular cloud cores, we find that
\beq
\rc \simeq 15 \left( \frac{\Omb}{10^{-14}\,{\rm rad\, s}^{-1}}\right)^2 \;
\left( \frac{30\, \muG}{B_{\rm ref}} \right)^2 \;
\left( \frac{m}{1\, M_{\odot}} \right) \: {\rm AU}. \label{rcn}
\eeq

The squared dependence on $\Omb$ and $\Bref$ means that $\rc$ can be quite
sensitive to the actual ambient cloud conditions.
The inverse dependence on $\Bref$ occurs since a stronger ambient magnetic
field implies a longer subcritical phase during which magnetic braking
can enforce $\Omega \simeq \Omb$. The relation (\ref{rc}) should be used
only when $\Bref$ is sufficiently strong that the core begins dynamic collapse
from a magnetically critical state rotating at a background rate $\Omb$.
One may also choose to write equation (\ref{rc}) in terms of
a critical column density $\sigma_{\rm crit}$ ($\simeq 2 \Bref/\mu_{\rm crit}$;
see \S\ 2.1) at which dynamical contraction begins.

\subsection{Self-Similarity and Dynamical Accretion from the Envelope}

\subsubsection{Analysis of Self-Similar Profiles}

The mass $m$ of the star-disk system grows by dynamical accretion from the 
nonequilibrium flattened envelope.
The accretion rate ($\dot{m}$) during the time $t>0$ 
determines how rapidly the star-disk system is built up.
Since the self-similar 
column density profile given by equation (\ref{sig}) is valid in at least
an innermost region, it is worthwhile to see what predictions
are made by self-similar solutions for the collapse from this profile.
We make comparisons with self-similar models and discuss below the likely 
evolution of the mass accretion rate for $t>0$ based on existing 
semianalytic and numerical models in the literature.

An important property of equation (\ref{sig}) is that it represents a
{\it nonequilibrium} column density profile, i.e., the column density exceeds
the equilibrium value. For comparison, the equilibrium
solution for a non-magnetic thin disk, the singular isothermal disk (SID), is
\beq
\sigma_{\rm SID} = \frac{\cs^2}{2 \pi G r} \label{sid}
\eeq
(see Basu 1997). This profile may be regarded as the thin-disk 
counterpart to the density profile of the equilibrium  
singular isothermal sphere (SIS), $\rho_{\rm SIS} = \cs^2/(2 \pi G r^2)$. 
The numerical results of BM94 yield a limiting profile which is overdense
relative to the SID by the factor $\ksid \simeq 2 \pi$. 
The factor $\ksid$ overdensity may be partially
accounted for by the presence of a magnetic field, as shown below. 

The magnetized solution can be recovered from the non-magnetic one in
thin-disk geometry by a simple scaling if the mass-to-flux ratio is spatially
uniform (Shu \& Li 1997; Nakamura \& Hanawa 1997; Basu 1997); the 
magnetic pressure force is then proportional to the thermal pressure force and
the magnetic tension is proportional (with opposite sign) to gravity.
Basu (1997) also shows that this scaling is only valid in an inner
region where the local magnetic field strength is much greater than the 
background field strength $\Bref$ of the ambient molecular cloud.
Under these circumstances, the hydrodynamic equations can be transformed
to the MHD equations by simply replacing $\cs^2$ with
\beq
c^{2}_{\rm s,eff} = (1 + \ampbyt)\, \cs^2, \label{scalecs}
\eeq
and $G$ with 
\beq
G_{\rm eff} = (1 + \amtbyg)\, G, \label{scaleg}
\eeq
where $\ampbyt$ is the ratio of
the accelerations due to magnetic pressure and thermal pressure, and
$\amtbyg$ is the ratio of the accelerations due to magnetic tension
and gravity. The three papers by Shu \& Li (1997), Nakamura \& Hanawa (1997)
and Basu (1997) all show that $\amtbyg = -\mu^{-2}$, where $\mu$ is the
mass-to-flux ratio in units of $\mu_{\rm crit} = (2 \pi G^{1/2})^{-1}$. 
However, they find slightly different
values for $\ampbyt$, although they are all essentially proportional
to $\mu^{-2}$. The proportionality constant can vary due to different 
assumptions about the vertical structure of the cloud or inclusion 
in the magnetic pressure 
of finite-thickness corrections to the magnetic tension force, as pointed
out by Shu \& Li (1997). For example, Nakamura \& Hanawa (1997) find that
$\ampbyt = \mu^{-2}$, but they do not include any finite thickness corrections
to the magnetic tension force. Basu (1997) finds that $\ampbyt = 2\,\mu^{-2}$,
since his magnetic pressure term actually incorporates a finite thickness
correction term from the magnetic tension. Since the magnetic pressure 
force is itself
an effect of the finite-thickness of the disk, we consider the inclusion
of the finite thickness magnetic tension term\footnote{The leading term in
the finite thickness correction to the magnetic tension is proportional to
$B_z^2 \nabla Z$, where $B_z$ is the vertical component of the magnetic
field, $Z$ is the half-thickness, and $\nabla$ is the gradient in the 
plane of the disk. This term exactly cancels
out one of the terms in the total magnetic pressure, proportional to
$-\nabla (Z\, B_z^2)$. The canceled term is exactly $-1/2$ times the total 
magnetic pressure in the late supercritical phase, so that the value of 
$\ampbyt$ calculated by Basu (1997)
is exactly twice that estimated by Nakamura \& Hanawa (1997).
A detailed derivation of the thin-disk equations that
were used by Basu (1997) can be found in CM93.}
a necessity. In the following
discussion, we scale hydrodynamic thin-disk solutions using 
$\ampbyt = 2\,\mu^{-2}$, both for this reason and also 
because we are comparing results with the simulations of BM94, for which the
analysis of Basu (1997) is most applicable.

Given that detailed numerical simulations of core formation and evolution
with ambipolar diffusion (e.g., FM93; CM94; BM94) show that $\mu\, (\simeq 2)$ 
has approximate spatial uniformity during the late stages
before protostar formation, we obtain the scaled magnetic equilibrium
solution, the singular isothermal magnetic disk (SIMD), from the SID
profile. Using the exact value $\mu = 2.1$ from the
late stages of BM94's standard model, we find that 
\beq
\sigma_{\rm SIMD}=\frac{(1 + 2\mu^{-2})}{(1 - \mu^{-2})}\, 
\frac{\cs^2}{2 \pi G r}
\; \simeq 1.88 \; \frac{\cs^2}{2 \pi G r}, \label{sigmag}
\eeq
so that the limiting profile of BM94 (eq. [\ref{sig}]) exceeds the 
magnetic equilibrium value by the
factor $\ksimd \simeq 2\,\pi (1/1.88) \simeq 3.3$\footnote{Centrifugal support 
is very negligible at
this stage (BM94; Basu 1997), and can be ignored when considering sources
of support.}. This feature of nonequilibrium column
density profiles, accompanied by supersonic infall velocities, is 
common to several independent models of magnetized cloud collapse
(FM93; CM94; BM94; Nakamura et al. 1995, 1998; Tomisaka 1996; Basu 1997),
and is reminiscent of the spherical similarity solution found by Larson (1969)
and Penston (1969) for the runaway collapse phase ($t<0$), in which the
density profile exceeds the SIS by the factor 4.4. This solution has
been extended to the accretion phase ($t>0$), when a central point mass
exists, by Hunter (1977); we subsequently refer to the
combined solution as the Larson-Penston-Hunter (LPH) solution. 
Given that there are an infinite number of spherical hydrodynamic 
similarity solutions, with the highly dynamic LPH solution and the
static (at $t=0$) Shu solution at two extremes (Hunter 1977; Whitworth \&
Summers 1985), it is remarkable that various numerical calculations
show a convergence to the LPH solution near the cloud center
(Larson 1969; Hunter 1977; Foster \& Chevalier 1993). The convergence
to the LPH solution has been justified recently by Hanawa \& Nakayama
(1997), who performed a stability analysis of various solutions.
Given this history in the hydrodynamic case, it is (at least
in retrospect) not
surprising that the MHD models also converge to a dynamic similarity solution.
However, the similarity solution in this case is quantitatively
different from LPH due to the distinctive character of self-gravity in 
thin-disk geometry. 

Basu (1997) used an approximate similarity solution (in which the 
self-similar profiles were predetermined by comparison with numerical
simulations) to analyze the approach to self-similarity during $t<0$
in a thin-disk magnetic cloud.
There has also been recent progress in finding exact solutions
to the self-similar equations for a thin disk.
Here, we utilize these solutions to gain further insight into the 
collapse properties both before and after protostar formation.
Li \& Shu (1997) found a similarity solution for $t>0$ which
starts from a static SIMD profile (as given by
eq. [\ref{sigmag}]) at $t=0$, and is similar in spirit to the spherical 
similarity solution of Shu (1977) for the collapse of a cloud that has a
static SIS profile at $t=0$. In contrast, Saigo \& Hanawa (1998, hereafter SH)
found hydrodynamic similarity solutions which exhibit dynamic collapse for 
$t<0$; they also extended these solutions to $t>0$. The magnetic solution
can be recovered from the hydrodynamic one using the scaling laws
given above. Their model also includes rotation, which leads to a
centrifugally-supported inner region when $t>0$, as we expect
based on the arguments given in \S\ 2.1. 
SH's solutions all converge to a spatially uniform infall velocity and
$r^{-1}$ column density profile at $t=0$, and for $t>0$ 
have an inner region with $r^{-1/2}$ profiles in infall velocity
and column density, both characteristic of free-fall. The rotating models
also have an innermost quasi-equilibrium region with zero
infall velocity and $r^{-1}$ column density profile. The latter 
profile for an equilibrium centrifugal disk can also be inferred from the
mass-radius relation of equation (\ref{rc}). SH's solutions 
are parametrized by varying values of the rotation parameter
$\omega = \cs j/Gm$. Using equations (\ref{sig}) and (\ref{jm}),
we find that the appropriate value for a rotating, magnetized core is 
\begin{eqnarray}
\omega & \simeq & \frac{\cs \Omb}{\Bref G^{1/2}} \nonumber \\
& = & 0.026 \left(\frac{\cs}{0.2 \,{\rm km \; s}^{-1}}\right)
\left(\frac{\Omb}{10^{-14} \,{\rm rad \; s}^{-1}}\right)
\left(\frac{30 \muG}{\Bref}\right). \label{omegamag}
\end{eqnarray}
SH have tabulated solutions for $\omega = 0$, 0.1,
and several higher values up to 0.495. 
Equation (\ref{omegamag}) shows that the typical rotating, magnetized core will
be an extremely slow rotator. It is best fit by SH's $\omega=0$ model,
since even their $\omega = 0.1$ solution agrees with the $\omega = 0$
solution to within a few percent in the
column density, infall velocity and mass accretion rate.
Although the nonrotating $\omega=0$ solution cannot give the inner centrifugal
disk size for $t>0$, this quantity can be obtained from equation (\ref{rc}).
The $\omega=0$ solution yields a column density $3.61 \, c^2_{\rm s,eff}/
(2 \pi G_{\rm eff} r)$ and infall velocity $-1.73 \, c_{\rm s,eff}$
at $t=0$, so that the scaled magnetic solution is
\begin{eqnarray}
\sigma & = & 
3.61 \, \frac{(1+2\mu^{-2})}{(1-\mu^{-2})} \frac{\cs^2}{2 \pi G r}
\simeq 6.78 \, \frac{\cs^2}{2 \pi G r}, \label{shcdens}\\
v_r & = & 
-1.73 \, (1 + 2\mu^{-2})^{1/2} \cs 
\simeq  -2.09 \, \cs . \label{shvel}
\end{eqnarray}
The latter term in each case represents the scaled magnetic solution
when $\mu = 2.1$.
The limiting profile of BM94 (eq. [\ref{sig}]) is within 10\% of the SH
value, but over 3 times the equilibrium value. Hence,
the inner regions of magnetized thin disks (where a similarity
solution is expected to be valid) appear to converge to the dynamic 
similarity solution of SH rather than the static (at $t=0$) solution of 
Li \& Shu (1997) during the approach to protostar formation
The infall velocity of equation (\ref{shvel}) agrees less well with the
simulations in regions where the column density is in good agreement, 
but Basu (1997) points out that the infall velocity begins
to converge to the self-similar value in a much smaller region
(see his Figures 3a and 4b and associated discussion
in \S\ 5.1.2). The maximum value of the infall velocity ($\simeq -\cs$) 
does not appear to have yet reached a terminal value in the BM94 model. 
This late convergence of the infall velocity, relative
to the column density, has also been noted by Nakamura et al. (1995, 1998)
In any case, a limiting infall velocity of $v_r \simeq -2 \cs$ is in 
agreement with the limiting ($t=0^{-}$) velocity estimated by Basu (1997) 
if the central 
magnetic force becomes significantly less than thermal-pressure force, and
is also in agreement with the limiting infall velocity obtained in the
thin-disk magnetic collapse calculations of Nakamura et al. (1995) and 
the two-dimensional magnetic simulation of Tomisaka (1996).

\subsubsection{Expected Mass Accretion Rates}

The late convergence of the infall velocity means that
only the collapse of the innermost mass shells can be described by the
similarity solution when extended to $t>0$; later mass shells will fall in 
at a lower rate since they start out with lower infall velocities at
$t=0$ (see further discussion in \S\ 5.1.3 of Basu 1997). 
A time-dependent accretion rate has been confirmed by Ciolek \& K\"onigl (1998)
for an initially subcritical magnetic cloud in which ambipolar diffusion
is operative. Additionally, Tomisaka (1996)
shows that an initially supercritical magnetic cloud with flux-freezing
also yields a time-dependent accretion rate. This feature was
also seen in the hydrodynamic collapse calculations of
Hunter (1977) and Foster \& Chevalier (1993). The likely outcome of a
time-dependent accretion rate has also been discussed recently by 
Henriksen, Andr\'{e}, \& Bontemps (1997), who present a simplified
hydrodynamic model for its development.

Given the reasonable fit of the SH solution, we estimate its predicted
mass accretion rate for a magnetic cloud.
SH find that $\dot{m}(t<0) = 6.27 \, c_{\rm s,eff}^3/G_{\rm eff}$ and
$\dot{m}(t>0) = 10.96 \, c_{\rm s,eff}^3/G_{\rm eff}$ in their $\omega = 0$
solution. Therefore, scaling to 
to the magnetic solution using $\mu = 2.1$ yields
\beq
\dot{m}(t=0^{-}) = 6.27 \, \frac{(1 + 2\mu^{-2})^{3/2}}{(1 - \mu^{-2})} \,
\frac{\cs^3}{G} \,
\simeq \, 14 \, \frac{\cs^3}{G} \label{mdotminus}
\eeq
just before protostar formation, and
\beq
\dot{m}(t=0^{+}) = 10.96 \, \frac{(1 + 2\mu^{-2})^{3/2}}{(1 - \mu^{-2})} \,
\frac{\cs^3}{G} \,
\simeq \, 25 \, \frac{\cs^3}{G} \label{mdotplus}
\eeq
just afterwards. The former value is in good agreement with
the limiting value $\dot{m}(t=0^{-}) \simeq 13 \, \cs^3/G$ estimated 
by Basu (1997), but the latter value
could not be obtained from his model which only applied to
$t<0$. These values can be compared with the values 
$\dot{m}(t=0^{-}) = 29 \, \cs^3/G$ and $\dot{m}(t=0^{+}) 
= 47 \, \cs^3/G$ of the spherical hydrodynamic LPH solution.

Equations (\ref{mdotminus}) and (\ref{mdotplus}) represent the limiting
forms for the inner (where $B_z \gg \Bref$) solution of BM94 {\it if} 
the contraction continues with
flux-freezing at a terminal value $\mu = 2.1$, and {\it also} if full 
convergence to the scaled SH similarity solution is achieved.  
For comparison, if the solution converged to the SIMD profile implied by
$\mu = 2.1$ at $t=0$, the mass accretion rates would be 
$\dot{m}(t=0^{-})=0$ and (using the model of Li \& Shu 1997)
$\dot{m}(t=0^{+}) = 1.05 (1 + H_0) \cs^3/G \simeq 2 \,\cs^3/G$, 
where $1 + H_0 = (1+2\,\mu^{-2})/(1-\mu^{-2}) \simeq 1.88$
is the overdensity factor supported by magnetic fields (see eq.[\ref{sigmag}]).
The $t=0^{-}$ limit of the BM94 simulation is better fit by
the scaled SH solution; the nonequilibrium column density is in good 
agreement and the magnitude of the infall velocity is far above zero,
although still a factor $\sim 2$ below the limiting SH value when the 
simulation ends. Despite the reasonable agreement of the 
BM94 simulation with the scaled SH solution for $t=0^{-}$, there is 
much important physics that is left out of a flux-frozen thin-disk 
similarity solution, as discussed below.

In reality, ambipolar diffusion plays an important role 
in the supercritical collapse
phase. Basu (1997) showed that the mass-to-flux ratio increases as
a weak power of the column density during $t < 0$:
\beq
\mu \propto \sigma^{\epsilon},
\eeq
where $\epsilon \simeq 0.05$ when using the simplified ionization model of
BM94. He also showed that this significantly reduces the level of
magnetic support $a_{\rm M}/|g_r| \propto \sigma^{-2\epsilon}$, so that 
the contraction of the initially near-equilibrium core becomes increasingly
dynamic.
However, the estimate of equation (\ref{mdotminus}), which assumes
constant $\mu$, is expected to be
reasonably good since much of the innermost $1\, M_{\sun}$ in the 
supercritical core is
in regions with $\mu \simeq 2$. When $t>0$, Ciolek \& K\"onigl (1998) have
shown that ambipolar diffusion has an even more dramatic effect; an increased
magnetic tension force in the vicinity of the newly formed protostar halts
the inward advection of magnetic field lines and ions, even as the
neutral particles tend to continue their infall. This results in a 
hydromagnetic shock at a larger radius where the ions and neutrals become
better coupled (see also Contopoulos, Ciolek, \&  K\"onigl [1998] for a 
semianalytic treatment). Ciolek \& K\"onigl (1998) find that 
$\dot{m}(t=0^{+}) \simeq 6\,
\cs^3/G$. The discrepancy with the estimate on the far right hand side of
equation (\ref{mdotplus})
is accounted for by three factors: 1) the mean mass-to-flux ratio $\mu$
in their supercritical core is somewhat higher than in the model of BM94,
2) the infall velocity magnitude is $\sim \cs$
rather than $\sim 2\, \cs$ when the central zone is converted to a sink cell,
and 3) the retarding effect of the ambipolar diffusion induced hydromagnetic
shock. In regard to point 2) above, Ciolek \& K\"onigl (1998) do find that
the maximum value of the infall velocity is still increasing when a 
sink cell is introduced at a central density $n_{\rm n,c} \approx 10^{11}$
cm$^{-3}$, when non-isothermal effects are expected to become significant.
Hence, it has not yet reached a terminal value, as expected if full 
convergence to an isothermal similarity solution has occurred. 
This raises the interesting point that nature may not provide enough
dynamic range between cloud core densities and the formation of a dense
opaque core for the infall velocity to fully converge to that of 
the SH isothermal similarity solution.

The previous discussion highlights the important fact that 
increasingly better understanding of the innermost regions of star-forming 
cores will depend crucially on detailed understanding of the microphysics 
of the attachment of neutral particles to the magnetic field and also
of the conditions under which opacity effects become important and a
hydrostatic stellar core is formed. This will lead to increasingly more
accurate estimates of $\dot{m}$ at and near $t=0$.\footnote{A detailed
understanding of the extent of magnetic coupling in the innermost region 
will also shed light on the possible dynamical role of magnetic fields
in the centrifugal disk phase.}
Based on current thin-disk MHD numerical solutions, we conclude that the
solutions are converging to a dynamic thin-disk similarity solution 
(approximately a scaled SH solution), but that the departure from flux-freezing
has two effects: 1) it accelerates the collapse somewhat during $t<0$ by
increasing the mass-to-flux ratio $\mu$ (Basu 1997), and 2) 
it significantly decelerates infall during $t>0$ for mass shells that 
start collapse outside the hydromagnetic shock caused by 
rapid ambipolar diffusion in the innermost region (Ciolek \& K\"onigl 1998).

All in all, the detailed physics is too complicated to be described by 
a single similarity solution. The numerical simulations must guide
us in the right direction. It is also worth remembering that
the vertical equilibrium assumption of the thin-disk approximation 
is expected to break down in the vicinity of the
protostar, where dynamical infall takes place. Here, the mass accretion
will occur in the vertical direction as well, so that the thin-disk accretion
rates, including the simulated value of Ciolek \& K\"onigl (1998), must
be regarded as lower limits to the actual value. Indeed, Tomisaka (1996)
found a mass accretion rate $\dot{m}(t=0^{+}) \approx 40\, \cs^3/G$ in his
two-dimensional axisymmetric MHD calculation, very close to the LPH value.
This simulation does include vertical infall, and the 
initial cloud is magnetically supercritical, unlike the model of
Ciolek \& K\"onigl (1998). Furthermore, it does not include ambipolar diffusion,
which acts to significantly retard the accretion rate at this stage.
Altogether, we believe that the true accretion rate must lie somewhere
between the Ciolek \& K\"onigl (1998) and Tomisaka (1996) values, with
\beq
\dot{m}(t=0^{+}) \gtrsim 10 \, \frac{\cs^3}{G} 
\eeq
a safe assumption. In other words, one cannot pin down an exact value
for the constant $m_0(0)$ due to the multiple complex phenomena involved,
but all models imply that it is of order 10. Based on these models, we would
further surmise that it falls within the range $10-20$.

All the hydrodynamic and MHD models do show that $\dot{m}$ decreases 
with time during $t>0$. The lower rates are due to infall of mass shells
which were moving slowly at $t=0$, as discussed by Basu (1997). 
In the hydrodynamic case, Foster \& Chevalier (1993) show that the 
collapse of Bonnor-Ebert spheres which have a 
long outer tail of mass in a distribution matching the SIS yield a 
mass accretion rate which approaches the value $\approx \cs^3/G$ implied by 
the Shu (1977) solution. This is due to the outer mass shells having very low
infall velocities when the central point mass is formed. In the MHD case, 
a static (at $t=0$) self-similar model (i.e., that of Li \& Shu 1997) is 
less likely to predict with any certainty an accretion rate for the outer
core material. This is due to the non-constancy of the mass-to-flux ratio in 
the outer core and the important effect of the background magnetic field
$\Bref$ in that region (e.g., CM94; BM94; Basu 1997). Both effects are
not included in the  self-similar Li \& Shu (1997) model. These effects also
yield a density profile that is somewhat shallower than $r^{-2}$ in the
outer core. Nevertheless, one may be able treat an appropriate value
from the Li \& Shu (1997) model (applied to the outer core) as a 
lower limit to (but not the actual value of) the mass accretion rate 
from the outer core (see Tomisaka 1996; Ciolek \& K\"onigl 1998).
The mass accretion rate can actually decrease even further if and when the
core mass is exhausted. The mechanism for ultimately terminating the accretion,
whether due to a limited mass supply or the effect of outflows from the
YSO, remains an unsolved problem.

\section{Constraints on Disk Evolution}

A centrifugal disk that is formed in the manner described in \S\ 2, with
radius $\rc$ given by equation (\ref{rc}), will contain most of
its matter in the outer disk. For typical parameters, less than 1\% of the
mass will fall directly onto the star. Hence, star formation requires that
the bulk of the protostellar 
material accrete through an initially self-gravitating centrifugal disk.
In order to accrete onto the central star, most mass shells must 
shed the bulk of their angular momentum.

Despite significant uncertainties in determining disk masses, or even the 
central star mass, various determinations yield YSO disk masses
which are about an order of magnitude
below the stellar mass (see reviews by Beckwith \& Sargent 1993; Mundy 1997).
This implies that the phase of a massive disk around a YSO is relatively
short lived, and that an efficient process is available for transferring
most of the system mass onto the central star.

The accretion may be driven by angular momentum loss from the outer
disk via a disk wind (e.g., K\"onigl 1989; Pelletier \& Pudritz 1992).
Alternatively, if outflows are generated within the innermost part of the
disk, nearby or at the star-disk interface, then the outer mass shells must 
find some other means to shed their angular momentum. 
This can still be accomplished by internal redistribution of angular momentum
within the disk. In the remainder of this paper, we assume that the 
disk evolution is driven by internal torques; either viscous or 
gravitational. Gravitational torques may dominate in the early stages,
when the disk mass is greater than or comparable to the stellar mass.
These can be driven by nonaxisymmetric spiral instabilities 
(e.g., Cassen et al. 1981; Heemskerk, Papaloizou, \& Savonije 1992; 
Laughlin \& Bodenhiemer 1994; Laughlin \& Rozyczka 1996), 
which may, if the instability is severe enough, also result in the formation 
of a secondary star, as argued by Adams, Ruden \& Shu (1989). In contrast,
viscous effects may induce a more gentle but steady accretion which will
be the primary mover when the disk mass is less than the stellar mass.
In the rest of this paper, we focus on the case of single-star
formation. However, if disk instabilities lead to the formation of a close
binary, the general feature of disk expansion discussed below should also
be present in a circumbinary disk.

In either the gravitational or viscous case, the inward transport of mass 
also requires that some
mass move outward in order to carry the angular momentum of the system;
therefore, the disk radius continually increases. The end result 
is that the disk radius increases dramatically, even though it
contains progressively less mass in relation to the central star.
This evolutionary scenario is consistent with 
the general tendency of any disk of a given angular momentum to
achieve its lowest energy state when all of the mass is brought
to $r=0$ and an infinitesimal mass carries all of the angular momentum
at an infinite radius (Lynden-Bell \& Pringle 1974).
Clearly, the observed YSO-disk systems have evolved significantly in this
direction if they started out with the massive disks envisioned by collapse
calculations. The recent time-dependent calculations of Hartmann et al. (1998)
illustrate the radial expansion of viscously evolving disks.

We next proceed to derive an estimate for the disk radius $\rd$ after it has
undergone significant evolution due to internal torques.
We let $m = \ms + \md$, where $\ms$ is the mass of the star and $\md$ is
the mass of the disk. If $\ms \gg \md$ at this late stage, 
as suggested by observations, then
the Keplerian angular velocity in the disk is
\beq
\Omega(r) = \frac{(G \ms)^{1/2}}{r^{3/2}}.
\eeq
Consequently, if the disk of mass $\md$ contains essentially all the angular
momentum $J$, then its radius is 
\beq
\label{rd1}
\rd = \frac{\ell}{G \ms} \left( \frac{J}{\md} \right)^2,
\eeq
where $\ell$ is a constant of order unity which depends on the assumed
column density profile in the disk. For $\sigma \propto r^{-p}$ ($p < 2$),
we find that $\ell = [(5/2-p)/(2-p)]^2$. 
We make contact with our collapse model by integrating equation (\ref{jm})
to yield the expected angular momentum 
\beq
\label{totalJ}
J \simeq \frac{1}{2} \frac{\Omb G^{1/2}}{\Bref} \, m^2.
\eeq
Combining equations (\ref{rd1}) and (\ref{totalJ}), we find that
\beq
\label{rd2}
\rd \simeq \frac{\ell}{4} \left( \frac{\Omb}{\Bref} \right)^2 \,
\left( \frac{m}{\ms} \right) \,
\left( \frac{m}{\md} \right)^2 \, m.
\eeq
The factor $\ell/4$ will be of order unity for most plausible column density
distributions, and will exactly equal 1 if $\sigma \propto r^{-3/2}$, which
is the inferred profile for the protosolar nebula (Weidenschilling 1977).
In the following discussion we simply let $\ell/4 = 1$.
For the case $m \approx \ms \gg \md$, and using equation (\ref{rc}), we find
\beq
\label{rd3}
\rd \simeq \left( \frac{\Omb}{\Bref} \right)^2\, \left( \frac{m}{\md} \right)^2
m \simeq \rc \, \left( \frac{m}{\md} \right)^2.
\eeq
Equation (\ref{rd3}) illustrates the growth of the disk radius $\rd$ from the
initial centrifugal value $\rc$ as mass is transferred from the disk to the 
star. Normalizing this equation to standard values yields
\beq
\label{rd4}
\rd \simeq 1500 \left( \frac{\Omb}{10^{-14}\,{\rm rad\, s}^{-1}}\right)^2 \;
\left( \frac{30\, \muG}{B_{\rm ref}} \right)^2 \;
\left( \frac{m}{1\, M_{\odot}} \right) \;
\left( \frac{m/\md}{10} \right)^2 \: {\rm AU}. 
\eeq

Taken together, equations (\ref{rc}) and (\ref{rd3}) bracket the 
possible radii for circumstellar disks. Collapse of a molecular cloud
core tends to build up a centrifugal disk of size $\rc$, which then 
grows in size (even as it loses mass to the central star) towards the
size $\rd$. The latter value should be treated as an upper limit; processes
such as outflows will at least partially compromise the assumed conservation
of mass and angular momentum and yield smaller final disk sizes.

\section{Discussion}

The general evolutionary picture that we have presented, in which an
inner centrifugal disk is formed within a flattened
nonequilibrium envelope (the remnant supercritical core), 
is consistent with several millimeter wave
studies of protostellar environments. For example, Hayashi, Ohashi, 
\& Miyama (1993) detected dynamical infall in an outer
flattened envelope of radius $\sim 1400$ AU surrounding the T Tauri star
HL Tau. The star is also observed to have an inner (presumably 
centrifugally supported) disk of radius $\lesssim 100$ AU (e.g., Beckwith
et al. 1990; Lay et al. 1994; Mundy et al. 1996). Recent studies of
the environment around the embedded protostars L1551-IRS 5 (Ohashi et al.
1996) and L1527-IRAS 04368+2557 (Ohashi et al. 1997a) also reveal evidence 
for dynamical infall in an envelope of size $\sim 1000-2000$ AU. 
L1551-IRS 5 is a well studied object and contains an
inner disk of radius $\lesssim 100$ AU (Keene \& Masson 1990; Lay et al. 1994;
Looney, Mundy, \& Welch 1997; the latter argue that the disk contains a 
binary system). 
These observations, particularly of HL Tau, imply that
infall from the envelope continues even after the inner disk has undergone
significant evolution. Hence, the transition from a inner disk of radius $\rc$
to one with radius approaching $\rd$ may not be a simple two-step process;
the star can be accreting from the inner disk even as the disk is growing by 
dynamical infall from the envelope. 

The estimated typical centrifugal radius
$\sim 10$ AU in our model is a reassuringly low value if 
the observed low mass $\sim 100$ AU inner disks are believed to be the
evolved (i.e., expanded) 
counterparts of these initial states. Normally, dense cores
of density $\sim 10^{4}$ cm$^{-3}$ and rotation rate 
$\sim 10^{-14}$ rad s$^{-1}$ would, if they conserve angular momentum,
yield centrifugal disks of radius $\sim 100$ AU. Hence, the observational
constraints would mean that these disks could hardly expand during accretion.
However, magnetic braking enforces background
rotation rates until much higher typical densities ($10^5 - 10^6$ cm$^{-3}$),
yielding typical centrifugal radii $\sim 10$ AU.
We have shown that the
expansion of such initial disks will still lead to final configurations
of size $\sim 1000$ AU. This has also been elegantly demonstrated by the 
time-dependent
calculations of Hartmann et al. (1998). Therefore, an explanation for the
size of
most YSO systems\footnote{A counterexample is the probable main-sequence star
$\beta$ Pic (and others of its class) which {\it do} have disks of size 
$\sim 1000$ AU.}
and our own solar system requires a mechanism to limit the disk expansion.
Since the expansion is driven by the need
to carry the angular momentum of the system, it will not be so dramatic 
if outflows can indeed carry
away a significant amount of angular momentum from the outer disk, as
suggested by K\"{o}nigl (1989). Another possibility 
is the photoevaporation of the outer disks (Shu, Johnstone, \& Hollenbach
1993; Hollenbach et al. 1994). The latter process is likely
effective in the Orion Nebula (Johnstone, Hollenbach, \& Bally 1998),
where the influence of the massive star $\theta^1$ C may be truncating
the outer disks of low-mass stars. 

As shown in \S\ 2.1, the initial centrifugal radius $\rc$ a simple
consequence of the existing density and angular velocity profiles in
the vicinity of the protostar as it is formed. Hence, it is useful to
compare our value of $\rc$ with that obtained in other environments.
Terebey et al. (1984, hereafter TSC) have presented such a value under the
assumption that the core is rigidly rotating and has the density profile
of a singular isothermal sphere $\rho_{\rm SIS} = \cs^2/(2 \pi G r^2)$.
Magnetic braking is invoked to justify the rigid rotation at a background
value $\Omb$ at this late stage. In this case, TSC show
that
\beq
\label{rcTSC}
\rc = \frac{\Omb^2 G^3 m^3}{16 \cs^8}.
\eeq
This expression (even more than eq. [\ref{rc}]) is very sensitive to input
parameters, but a typical value is
\beq
\rc = 39\, \left( \frac{\Omb}{10^{-14}\,{\rm rad}\,{\rm s}^{-1}} \right)^2
\left( \frac{0.2\, {\rm km}\,{\rm s}^{-1}}{\cs} \right)^8
\left( \frac{m}{1\, M_{\sun}} \right)^3 \, {\rm AU}.
\eeq
Hence, both equations (\ref{rc}) and (\ref{rcTSC}) predict typical centrifugal 
radii $\sim 10$ AU using different profiles of physical variables at $t=0$.
However, equation (\ref{rc}) incorporates the magnetic field and is based
on the result of a detailed calculation of the collapse phase ($t < 0$). 
BM94 have shown that
magnetic braking becomes relatively ineffective when a core becomes
supercritical, so that a period of angular momentum conservation precedes
the formation of a central protostar; differential rotation is established
in the core during this time. The column density also exceeds the 
equilibrium value, as discussed in \S\ 2. 

An important difference between equations (\ref{rc}) and (\ref{rcTSC})
is the {\it rate} at which the centrifugal disk is built up.
The TSC disk is built up with radius proportional to $m^3$, whereas
the centrifugal disk within an infalling magnetic
envelope is built up with radius proportional to $m$. 
A relation very similar to equation (\ref{rcTSC}), differing only by a
multiplicative constant, would apply to the magnetized disk as well
if it were rigidly rotating at $t=0$.
Figure 1 illustrates the growth of $\rc$ versus mass in both cases,
using typical molecular cloud parameters. In the magnetic case, the
disk grows more rapidly with mass early on, but eventually lags in 
growth rate due to the $m^3$ growth of the TSC disk. 
The growth of mass versus time (the mass accretion rate) also differs
in the two models, as discussed in \S\ 2. The TSC model assumes self-similarity
for all mass shells, hence $\dot{m} = m_0 \cs^3/G$ with $m_0=0.975$,
as implied by the self-similar collapse model for a singular isothermal
sphere (Shu 1977). The centrifugal disk in a realistic magnetized cloud grows
due to a mass accretion rate $\dot{m} = m_0(t) \cs^3/G$, where $m_0(0)$
is conservatively estimated here to be $\approx 10$ (see \S\ 2.2.2).
Since $m_0(t)$ will gradually decrease with time when
$t>0$, we can only use $m_0 \approx 10$ at early times.

This early phase of protostellar evolution, when the two values of $\rc$ 
differ the most, can be
associated with the Class 0 phase (Andr\'{e}, Ward-Thompson, \& Barsony 1993). 
The age of these
protostars is inferred to be only up to a few $\times \, 10^4$ yr.
Circumstellar disks are not easily detected in this highly embedded phase. 
However, Pudritz et al. (1996) find indirect evidence for a disk around
the prototypical Class 0 source VLA 1623 and place an upper limit to its
radius of 175 AU. It is not clear what fraction of the flattened structure
is a centrifugal disk.
Like most Class 0 sources, VLA 1623 drives a powerful 
outflow. In fact, Class 0 objects tend to have the most powerful CO outflows 
(Bontemps et al. 1996; Bachiller 1996).
If outflows are in fact powered by the presence of a centrifugal disk, 
we should expect that
significant (though deeply embedded) disks exist in this early phase.
However, a model that assumes rigid rotation at $t=0$ predicts a very slow
early growth of a disk. For example, using the TSC model 
with $t=3 \times 10^4$ yr, $\cs = 0.2$ km s$^{-1}$, and
$\Omb = 10^{-14}$ rad s$^{-1}$ yields a radius $\rc = 6.5 \times 10^{-3}$ AU
$ = 1.38\, R_{\sun}$. This is less than a typical protostellar radius
of $\approx 1.5 - 3 \, R_{\sun}$ (Stahler 1988), so that infall is 
still occurring directly onto the protostar; there is no centrifugal disk.
The magnetic model presented here, with differential rotation (therefore
higher rotation rate near $r=0$) and a higher
initial infall rate, in which we conservatively choose $m_0(0) = 10$,
yields $\rc = 8.4$ AU at the same time. Aside from the additional parameter
$\Bref = 30\, \muG$, all other parameters are the same as in the previous case.
Hence, in this model, a centrifugal disk is predicted 
to exist in the early protostellar phase given standard molecular cloud
input parameters.

\section{Summary}

Using the results of earlier numerical and semianalytic models, we have
found a connection between ambient conditions in a magnetized molecular 
cloud and the
the size of disks around YSO's. Our analytic expression (eq. [\ref{rc}])
relates the centrifugal radius to the mass via two fundamental
parameters characterizing the ambient molecular cloud:
the background rotation rate $\Omb$ and a background magnetic field
strength $\Bref$. This represents an estimate for the initial disk radius
formed from the collapse of a molecular cloud core. An upper limit to 
the disk radius after it has undergone significant evolution due to internal 
torques is given by equation (\ref{rd3}); it is simply the initial
value scaled by the square of the total system mass to the disk mass
$(m/\md)^2$ ($\approx 100$ according to observations). 

We have reviewed a variety of numerical and semianalytic models which
support the general view that a centrifugal disk is built up from
an infalling evelope which begins dynamic collapse before a central
protostar is formed at time $t=0$. This leads to a mass accretion rate 
$\dot{m} = m_0(t) \cs^3/G$, where $m_0$ is likely $\gtrsim 10$
at $t=0$, but subsequently
decreases with time. This is due to the fact that convergence
to self-similarity (especially in the infall velocity) takes place
only in an innermost region near $r=0$ and $t=0$. The density profiles
converge to self-similar values over a larger range. The inner
regions of thin-disk MHD
simulations are shown to be in good agreement (at least during $t < 0$) with 
hydrodynamic thin-disk similarity solutions found by Saigo \& Hanawa
(1998), when the latter are scaled in an appropriate manner.

We conclude that the detection of centrifugal disks around very young 
($\sim 10^4$ yr) Class 0 protostars will support the view that a cloud core
begins dynamic collapse and develops differential rotation before
a central protostar is formed. The consequent high rotation rates in the
deep interior and the rapid infall mean that an inner centrifugal disk
is initially built up more rapidly than in equilibrium, rigidly rotating 
cores. Therefore, a significant centrifugal disk is present in 
the early Class 0 stage and can help to drive the strong outflows from
these sources.

Our model places constraints on the size of circumstellar disks in the
early and late stages of evolution, regardless of the internal processes 
which may drive their evolution. These constraints should aid interpretation
of observations and act to sharpen our
intuition for future numerical simulations of the accretion phase in
rotating, magnetized cloud cores.

\acknowledgements

I thank Glenn Ciolek for many illuminating discussions. 
This work was supported by the Natural Sciences and Engineering
Research Council of Canada.

\clearpage

\begin{figure}
\plotone{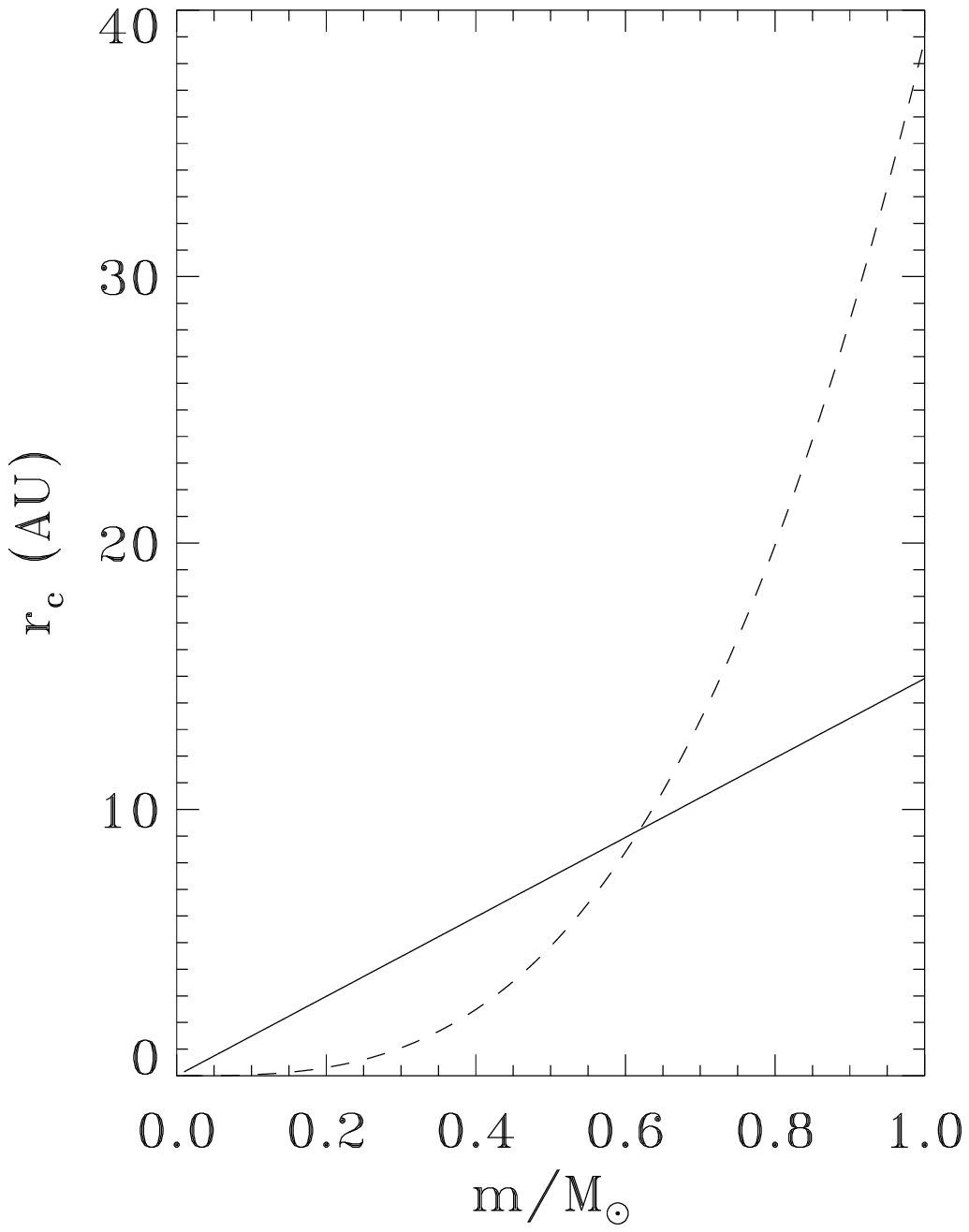}
\caption{
A comparison of the growth of centrifugal
radius $\rc$ versus mass $m$ for two cases. Solid line: a magnetized
core with column density given by equation (\ref{sig}) and differential
rotation given by equation (\ref{omega}) at $t=0$. 
Dashed line: a singular isothermal
sphere that is rigidly rotating at $t=0$. 
Standard input parameters $\cs = 0.2$ km s$^{-1}$ and $\Omb = 10^{-14}$
rad s$^{-1}$ are used in both cases, and $\Bref = 30\, \muG$ is also used
for the magnetized case. 
}
\end{figure}

\end{document}